\documentclass[preprint,showpacs,preprintnumbers,amsmath,amssymb]{revtex4}
\usepackage{hyperref}
\usepackage{graphics}
\usepackage{graphicx}
\usepackage{dcolumn}
\usepackage{bm}
\usepackage[latin1]{inputenc}
\usepackage[english]{babel}
\usepackage{amsmath}

\begin{document}

\preprint{}

\title{Quantum Behavior of Measurement Apparatus\\}

\author{Taoufik Amri\footnote{Corresponding author : amri.taoufik@gmail.com}}

\affiliation{\begin{small}Laboratoire Kastler Brossel, UMR 8552 (ENS/UPMC/CNRS), 4 Place Jussieu Case 74, 75252 Paris Cedex 05, France\end{small}}

\date{\today}

\begin{abstract}
We precise for the first time the quantum behavior of a measurement apparatus in the framework of the usual interpretation of quantum physics. We show how such a behavior can also be studied by the retrodiction of pre-measurement states corresponding to its responses.
We translate in terms of these states some interesting properties of the behavior of an apparatus, such as the projectivity, the fidelity, the non-Gaussian character, or the non-classicality of measurements performed by this one.
We also propose an experimental procedure allowing the tomography of these pre-measurement states for optical detectors. 
We illustrate the relevance of these new notions for measurements, by evaluating them for two detectors widely used in quantum optics: the avalanche photodiode and the homodyne detection.
\end{abstract}


\maketitle

\section{Introduction}
The measurement apparatus plays an important role not only in quantum physics experiments, by providing information about the measured system, but also in the foundations of quantum theory by leading to the famous \textit{measurement problem} \cite{Neumann1955,Zurek1982}.
This one is in part linked to our ability to prepare the measured system in a particular state, by using information available after its interaction with an apparatus. 
This state, conditioned on the measurement result, is given in the simplest case by Von Neumann's \textit{projection rule} \cite{Neumann1955}. 
In the most general case, this is a non-linear transformation of the initial state that we will specify in the following.

However, such a conditioning needs a precise knowledge about the performed measurements. It is surprising that the first experimental characterization of measurements has been realized only very recently, in quantum optics \cite{QDT1}, while the quantum tomography of states (QST) and processes (QPT) \cite{QPT} are now well-established in this same field.
Indeed, the quantum detector tomography (QDT) is the reconstruction of positive operator valued measures (POVMs) \cite{Nielsen2000}, describing any measurement device. We probe the behavior of its responses with a set of known states, in order to reconstruct the POVM elements giving the probabilities which are the closest to those measured. This was proposed for instance in Ref. \cite{QDT2} with a maximum-likelihood estimation (MaxLik), which is a reconstruction method widely used for QSTs.

Despite attempts, the interpretation of QDT results is not yet exploited, and to our knowledge, it has never been treated in the framework of the usual interpretation of quantum physics, which mainly deals with states in its formalism.
The aim of this paper is to show that the retrodictive approach of quantum physics \cite{Aharonov1964, Barnett2000}, in addition to the usual predictive one, provides unexpected insights into \textit{the quantum behavior of a measurement apparatus}. 
It brings a meaning to \textit{the non-classicality of a measurement} and allows us to introduce other interesting properties for a measurement, such as its \textit{projectivity}, its \textit{fidelity with another measurement}, and its \textit{non-Gaussian character}. Moreover, we evaluate some of these properties for two detectors widely used in quantum optics: the avalanche photodiode and the homodyne detection.

\section{States and propositions}
\label{state_proposition}
As a preliminary step, it is necessary to focus on some important notions from the usual interpretation of quantum physics, which is predictive about measurement results and retrodictive about state preparations.
These tasks need conditional probabilities linked by Bayes' theorem, and in the most general case, the expression of probabilities on the Hilbert space $\mathcal{H}$ is given by the recent generalization (2003) \cite{Bush2003} of Gleason's theorem \cite{Gleason1957}.

This theorem is only based on very general requirements about the probabilities and the mathematical structure of the Hilbert space, which is the starting point of any quantum description.
It states that any system can be described by a \textit{density operator}. 
More precisely, when we propose \textit{a property }$P_{n}$ about a system corresponding to \textit{a precise value for a given observable}, the proposition about the state of this system is represented by the projector on the eigenstates corresponding to this value.
In the most general case, such \textit{a proposition} can also be represented by \textit{a hermitian and positive operator} $\hat{P}_{n}$ in the Hilbert space. 
The probability $\mathrm{Pr}\left(n\right)$ of checking this property on the system should satisfy the following conditions: 
\begin{enumerate}
\item $0 \leq\mathrm{Pr}\left(n\right)\leq 1$ for any proposition $P_{n}$.
\item $\sum_{n}\,\mathrm{Pr}\left(n\right)=1$ for any \textit{exhaustive} set of propostions such that $\sum_{n}\,\hat{P}_{n}=\hat{1}$.
\item $\mathrm{Pr}\left(n_{1}\,\textrm{or}\,n_{2}\,\textrm{or}\,...\right)=\mathrm{Pr}\left(n_{1}\right)+\mathrm{Pr}\left(n_{2}\right)+...$ for any \textit{non-exhaustive} set of propositions such that $\hat{P}_{n_{1}}+\hat{P}_{n_{2}}+...\leq \hat{1}.$
\end{enumerate}

According this theorem for a system needing predictions (i.e. with a Hilbert space of dimension $D\geq 2$), this probability is given by $\mathrm{Pr}\left(n\right)=\mathrm{Tr}\lbrace\hat{\rho}\,\hat{P}_{n}\rbrace$ in which $\hat{\rho}$ is a \textit{hermitian, positive,} and \textit{normalized} operator, allowing us to make predictions about any properties of the system.
This is the reason why we call this operator \textit{the state of the system}, and the probabilities are in fact conditioned on this state of knowledge.
When we reduce this operator in its eigenbasis, we retrieve a statistical mixture usually used in teaching for introducing the density matrix.
Thus, each time that we can write the probabilities concerning any \textit{exhaustive} set of propositions about the system in this previous form, we can then determine the state of the system thanks to this theorem.

\section{Preparations and Measurements}
\label{pm}
\subsection{The Game}
\begin{figure}[h!]
\begin{center}
\includegraphics[scale=0.35]{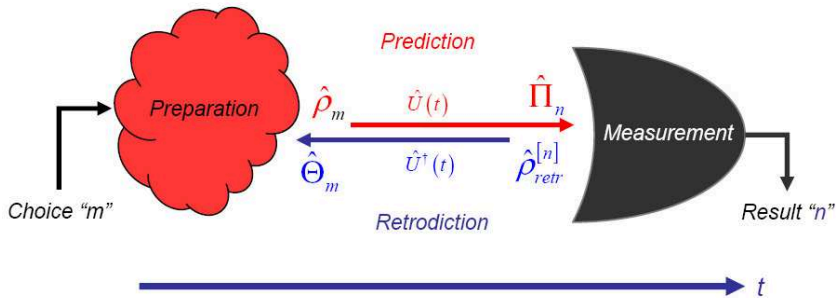} 
\end{center}
\caption{\label{figure1} (color online) The game of preparations and measurements in quantum physics: \textit{the prediction and the retrodiction need quantum states and propositions that we respectively note $\left(\hat{\rho}_{m}, \hat{\Pi}_{n}\right)$ and $\left(\hat{\rho}_{\mathrm{retr}}^{[n]}, \hat{\Theta}_{m}\right)$. The time-evolution operator $\hat{\mathcal{U}}\left(t\right)$ allows us to propagate forward or backward in time these states between the preparation and the measurement, in order to make these predictions.}}
\end{figure}

In quantum physics, any situation is based on preparations and measurements. 
In such a game, the preparation of the system in a state $\hat{\rho}_{m}$ can be associated to a classical information that we call the \textit{choice} 'm'. 
The measurement, corresponding to the POVM element $\hat{\Pi}_{n}$, gives another classical information which is simply the \textit{result} 'n'.
However, we can only make predictions about these choices 'm' and these results 'n'. We have then two approaches in quantum physics, that we will examine in details in the following.
Each approach needs a quantum state and propositions, allowing predictions about the measurement results or retrodictions about the state preparations, as pictured on Fig. \ref{figure1}.
 
\subsection{Predictive approach}
\label{predictive}
We usually prepare the system in a given quantum state based on a choice 'm', and we make propositions about the results of any subsequent measurement which will be performed on this system.
The conditional probability of obtaining a certain result 'n', after that the system was prepared in the state $\hat{\rho}_{m}$, is then given by:
\begin{equation}
\label{conditional_proba}
\mathrm{Pr}\left(n\vert m\right)=\mathrm{Tr}\lbrace\hat{\rho}_{m}\hat{\Pi}_{n}\rbrace .
\end{equation}
This result constitutes the Born's rule, in which the proposition operator $\hat{\Pi}_{n}$ is nothing else than the POVM element corresponding to the result 'n'. 
In the framework of states and propositions, the POVM elements should be understood as propositions about the state of the system interacting with the measuring device. 
These operators also describe the behavior of responses of this measuring device, and the corresponding propositions are simply labeled by the responses of the apparatus performing their tests.

Moreover, any proposition can be reduced to more \textit{simple propositions} corresponding to \textit{projectors} on a larger Hilbert space. 
This is the Neumark's extension \cite{Neumark1983} which can often be interpreted as a noise influence on ideal measurements.
Such a modeling is in fact common in quantum optics. An imperfect optical measurement can be viewed as an ideal measurement onto the signal field, which has interacted with an appropriate noise field.

\subsection{Retrodictive approach}
\label{retrodictive}
Contrary to the previous one, this approach is less usual and leads to difficulties linked to an inappropriate use of its tools (see for example \cite{Aharonov1995}).
We start with an example in order to "demystify" such a approach.

We consider a perfect photon counter able to discern the number of absorbed photons. 
When such a device displays $n$ counts, the proposition about the measured system simply corresponds to the projector $\hat{P}_{n}=\vert n\rangle\langle n\vert$. 
The pre-measurement state, which is sufficient to describe this result, is then this photon number state $\hat{P}_{n}$. 
This one should allow us to predict the states in which the measured system was prepared, before its interaction with the apparatus giving the result on which we base such a retrodiction.

The link between the proposition checked by the apparatus and the pre-measurement state is obvious in this ideal case. We will generalize it by using the generalization of Gleason's theorem [\ref{state_proposition}]. Indeed, we can already write an expression for the retrodictive probability. 
The probability of preparing the measured system in a given state $\hat{\rho}_{m}$, when we have the result 'n', can be written as
\begin{equation}
\label{retrodictive_proba}
\mathrm{Pr}\left(m\vert n\right)=\mathrm{Tr}\lbrace\hat{\rho}_{\mathrm{retr}}^{[n]}\hat{\Theta}_{m}\rbrace.
\end{equation}
In this expression, $\hat{\rho}_{\mathrm{retr}}^{[n]}$ is the pre-measurement state retrodicted from the result 'n' labeling the POVM element $\hat{\Pi}_{n}$. 
$\hat{\Theta}_{m}$ is a hermitian and positive operator, corresponding to a proposition about the state of the measured system just after the process preparing the state $\hat{\rho}_{m}$. 
In order to have an \textit{exhaustive} set of propositions about the preparations of the system, these operators should constitute a resolution of the Hilbert space $\mathcal{H}$:
\begin{equation}
\label{normalization_preparation}
\sum_{m}\,\hat{\Theta}_{m}=\hat{1}.
\end{equation}
As it was previously noticed by S. Barnett et al. \cite{Barnett2000}, the expressions of the retrodicted state and proposition operators can be derived from Born's rule with Bayes' theorem.
However, it is interesting to give them in the light of the generalization of Gleason's theorem, which justifies the expression of retrodictive probabilities and brings interesting insights into these tools.

Retrodiction requires no additionnal assumption other than the projection rule which could be used in the preparation of states. Indeed, Bayes' theorem gives for the retrodictive probability: 
\begin{equation}
\label{retro_proba}
\mathrm{Pr}\left(m\vert n\right) = \mathrm{Pr}\left(n\vert m\right)\mathrm{Pr(m)}/\mathrm{Pr}(n).
\end{equation}
The marginal probability $\mathrm{Pr}\left(n\right)$ of having the result 'n' is obtained by summing the joint probability on all 'candidate' states: 
\begin{equation}
\label{marginal_proba}
\mathrm{Pr}\left(n\right)=\sum_{m}\,\mathrm{Pr}\left(n\vert m\right)\mathrm{Pr(m)}= \mathrm{Tr}\lbrace\hat{\rho}^{[?]}\hat{\Pi}_{n} \rbrace,
\end{equation}
in which we introduce the state $\hat{\rho}^{[?]}=\sum_{m}\,\mathcal{P}_{m}\,\hat{\rho}_{m}$.
In a conditional preparation, such a state corresponds to the state of the measured system after an 'unread' measurement \cite{Haroche2006}: a mixture of states conditioned on each result 'm' and weighted by their respective success probabilities $\mathcal{P}_{m}=\mathrm{Pr(m)}$. 
This mixture could also be obtained with preparations based on random choices 'm', as we will see it in our scheme for the tomography of pre-measurement states.

Then, for writing the retrodictive probabilities (\ref{retro_proba}) in their most general form (\ref{retrodictive_proba}), this statistical mixture should be such that $\hat{\rho}^{[?]}=\hat{1}/D$ in a Hilbert space $\mathcal{H}$ of finite dimension $D$.
The pre-measurement state simply corresponds to the normalized POVM element:
\begin{equation}
\label{retro_state}
\hat{\rho}_{\mathrm{retr}}^{[n]}=\frac{\hat{\Pi}_{n}}{\mathrm{Tr}\lbrace\hat{\Pi}_{n}\rbrace},
\end{equation}
and the proposition operators are given by:
\begin{equation}
\label{proposition_operator}
\hat{\Theta}_{m}=D\mathcal{P}_{m}\hat{\rho}_{m}.
\end{equation}
Their resolution of the Hilbert space (\ref{normalization_preparation}) is in fact equivalent to the maximization of Von Neumann entropy $\mathcal{S}\left[\hat{\rho}\right]=-\mathrm{Tr}\lbrace\hat{\rho}\log{\hat{\rho}}\rbrace$ by the mixture $\hat{\rho}^{[?]}$. This one is then maximally mixed and probes all the responses of the apparatus.

\begin{figure}[!]
\begin{center}
\includegraphics[scale=0.35]{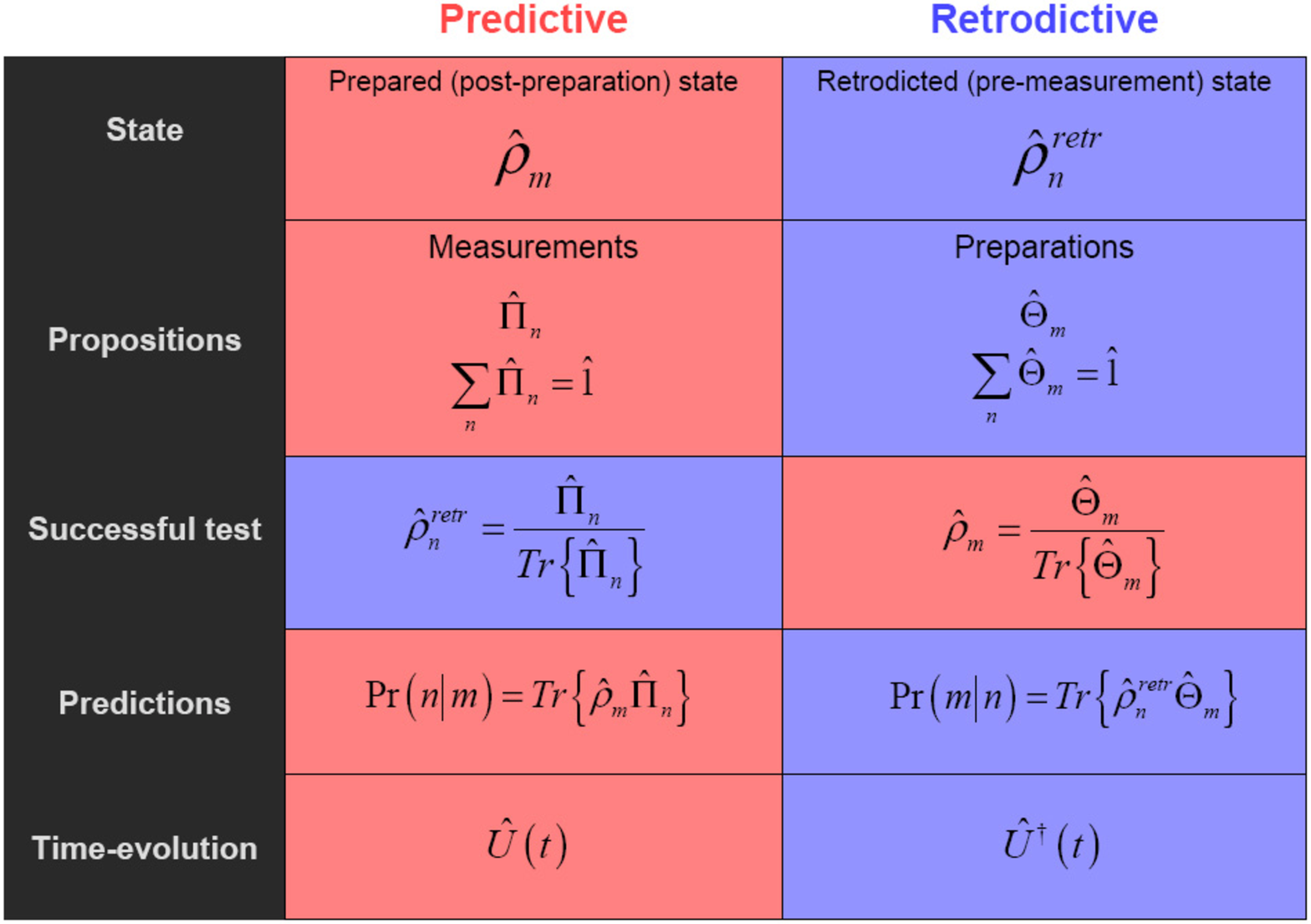} 
\end{center}
\caption{\label{figure6} (color online) Complementarity between the predictive and retrodictive approaches of quantum physics -  \textit{the successful test of a proposition in one of these approaches gives the state of the other one, with which we can make predictions about measurement results or preparation choices.}}
\end{figure}

Finally, we propagate these retrodicted states backward in time contrary to prepared states, as pictured on Fig. \ref{figure1}. Indeed, when we take the Heisenberg picture for the POVM elements in the predictive approach, we retrieve the pre-measurement state (\ref{retro_state}) in the Schrödinger picture with the time-evolution operator $\hat{\mathcal{U}}\left(-t\right)=\hat{\mathcal{U}}^{\dagger}\left(t\right)$.
The complementarity between the predictive and retrodictive approaches is illustrated on Fig. \ref{figure6}.

\section{How to study an apparatus ?} 
The POVM elements should be understood as propositions about the measured system. These propositions are actually labeled by the responses of the measuring device, and this one appears as a physical implementation of tests checking these propositions.
However, the POVM elements only allow us to study the behavior of an appartus, by giving the predictive probabilities (\ref{conditional_proba}) of having a certain response. Thus, how could we go beyond these simple conditional probabilities in order to study the measurement apparatus ?

\subsection{Measurement effects on a system}
\label{measurement_effects}
The POVM elements are not sufficient for studying such effects in the general case. 
Indeed, the effects of a \textit{generalized measurement} on a state are given by the following transformation (see for example \cite{Hornberger2009}):
\begin{equation}
\label{effects_measurement}
\hat{\rho}\rightarrow\hat{\rho}_{\mathrm{cond}}^{[n]}=\sum_{\mu}\frac{\hat{M}_{n,\mu}\,\hat{\rho}\,\hat{M}_{n,\mu}^{\dagger}}{\mathcal{P}_{n}},
\end{equation}
where $\hat{M}_{n,\mu}$ are the measurement operators, also called Kraus operators \cite{Kraus1983}, describing completely the measurement process corresponding to the result 'n'. 
The probability $\mathcal{P}_{n}$ of having this result is then given by the POVM element:
\begin{equation}
\label{POVM_decomposition}
\hat{\Pi}_{n}=\sum_{\mu}\,\hat{M}_{n,\mu}^{\dagger}\hat{M}_{n,\mu}.
\end{equation}
The normalization of expression (\ref{effects_measurement}) by the probability $\mathcal{P}_{n}$ is the manifestation of the so-called \textit{projection rule}, leading to a non-linear transformation of the initial state contrary to an unread measurement.
When we perform the measurement on only one part B of a bipartite entangled resource $\hat{\rho}_{AB}$, the state of the other part A - conditioned on the result of this measurement - depends on the POVM element corresponding to the expected result 'n':
\begin{equation} 
\label{cond_state}
\hat{\rho}_{\mathrm{A, cond}}^{[n]}=\frac{1}{\mathcal{P}_{n}}\,\mathrm{Tr}_{B}\lbrace\hat{\rho}_{AB}\,\hat{1}_{A}\otimes\hat{\Pi}_{n}\rbrace.
\end{equation}
In quantum optics, such protocols have been recently implemented for preparing more "exotic" non-classical states from gaussian resources, such as Fock states \cite{Lvovsky2001} or Schrödinger's cat states \cite{Ourjoumtsev2006}.

Thus, in the predictive approach [\ref{predictive}], the study of an apparatus is restricted to non-destructive measurements. Indeed, it needs a kind of QPT for determining the measurement operators $\hat{M}_{n,\mu}$. 
Such a approach also depends on the initial state of the measured system, and cannot lead to a study of the apparatus as an implementation of tests checking propositions.
Furthermore, different measurement processes $\lbrace\hat{M}_{n,\mu}\rbrace_{\mu}$ can check one same proposition, since the decomposition of a given POVM element (\ref{POVM_decomposition}) is not unique.
In other words, there is no general rule to design a measuring device with a given POVM, underlying the importance of an experimental determination of POVM elements describing an apparatus \cite{QDT1}.
Therefore, the predictive approach is more adapted for studying the measurement effects on a system, instead to really translate its behavior in terms of states.

\subsection{'State' translation of measurements}
\begin{figure}[h!]
\begin{center}
\includegraphics[scale=0.3]{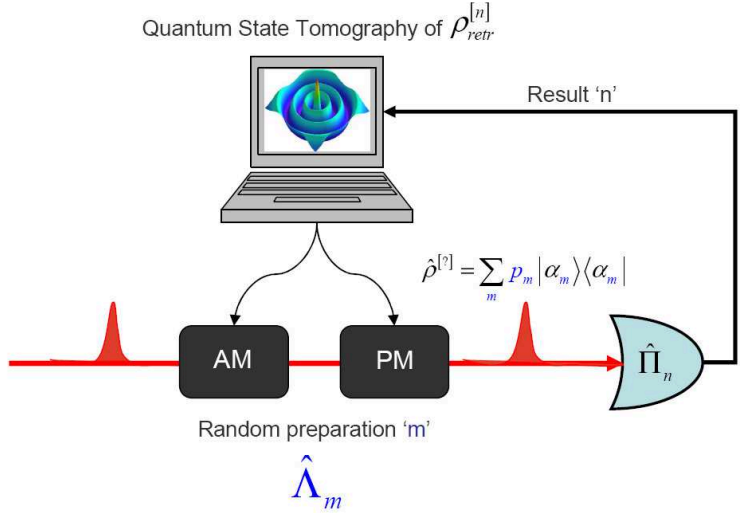} 
\end{center}
\caption{\label{figure2} (color online) Scheme for the quantum tomography of pre-measurement states for optical detectors: Each light pulse is randomly prepared in a coherent state $\vert\alpha_{m}\rangle$, with a probability $\mathcal{P}_{m}$, by an amplitude modulation (AM) and a phase modulation (PM).}
\end{figure}

The retrodictive approach [\ref{retrodictive}] seems to be more close to such a goal. 
Indeed, from a given result, we make predictions about the preparations of the system before its interaction with the apparatus checking the property corresponding to this result. 
The main tool of this approach is the retrodicted pre-measurement state, that we propagate backward in time for predicting these preparations as depicted on Fig. \ref{figure1}. We can therefore determine in which kind of states the system was prepared for leading to such a result.
Actually, the pre-measurement states of an apparatus can directly be reconstructed with the same tools as a QST. This experimental procedure is pictured on Fig. \ref{figure2} for optical detectors.

In such a experiment, we can probe the behavior of an optical detector with a statistical mixture of coherent states $\hat{\rho}^{[?]}=\sum_{m}\,p_{m}\vert\alpha_{m}\rangle\langle\alpha_{m}\vert$, obtained by amplitude and phase modulations randomly performed on each light pulse.
As a first step, the QST of this mixture could be necessary and leads to its Cholesky decomposition, $\hat{\rho}^{[?]}=\hat{\sigma}^{\dagger}\hat{\sigma}$.
We must have $\mathrm{det}\lbrace\hat{\rho}^{[?]}\rbrace\neq 0$ for generating the proposition operators:
\begin{equation}
\label{exp_proposition_preparation}
\hat{\Lambda}_{m}=\left(\hat{\sigma}^{-1}\right)^{\dagger}\,p_{m}\vert\alpha_{m}\rangle\langle\alpha_{m}\vert\,\hat{\sigma}^{-1}.
\end{equation} 
With these operators, we realize an usual QST of the state $\hat{\rho}_{n}$ giving the conditional probabilities $\mathrm{Pr}\left(m\vert n\right)=\mathrm{Tr}\lbrace\hat{\rho}_{n}\hat{\Lambda}_{m}\rbrace$, measured directly in the experiment.
For this purpose, we replace the POVM elements describing the measurements in a QST method by these proposition operators (\ref{exp_proposition_preparation}). We determine then the state $\hat{\rho}_{n}$ giving the closest probabilities $\mathrm{Pr}\left(m\vert n\right)$ to those measured. 
The pre-measurement state, retrodiced from the response 'n' of the apparatus, is obtained from this state by:
\begin{equation}
\hat{\rho}_{\mathrm{retr}}^{[n]}=\frac{\hat{\sigma}^{-1}\hat{\rho}_{n}\left(\hat{\sigma}^{-1}\right)^{\dagger}}{\mathrm{Tr}\lbrace\hat{\sigma}^{-1}\hat{\rho}_{n}\left(\hat{\sigma}^{-1}\right)^{\dagger}\rbrace}.
\end{equation}

Finally, it's interesting to note that the retrodictive probabilities (\ref{retro_proba}) can also be obtained from data taken in QDT experiments \cite{QDT1}, in which we directly measure the predictive probabilities (\ref{conditional_proba}) for each response 'n' of an apparatus.
Indeed, if the preparation rate is the same for all the probe states $\hat{\rho}_{m}$, the probability of preparing the state $\hat{\rho}_{m}$ is simply given by $\mathrm{Pr}\left(m\right)=1/M$, where $M$ is the number of probe states.
These retrodicted probabilities are then given by:
\begin{equation}
\label{retro_probaQDT}
\mathrm{Pr}\left(m\vert n\right)=\frac{\mathrm{Pr}\left(n\vert m\right)}{\sum_{m'=1}^{M}\,\mathrm{Pr}\left(n\vert m'\right)}.
\end{equation}
We can therefore realize the QST of pre-measurement states, retrodicted from each response of the device, with the procedure previously depicted.

\section{Behavior of a measurement apparatus}
\label{behavior_apparatus}
We translate some interesting aspects of such a behavior in terms of pre-measurement states. 
This translation motivates the implementation of our experimental procedure for the tomography of these states, or the interpretation of results from realized experiments \cite{QDT1}.
We also illustrate the relevance of these new properties for measurements on two well-known detectors from quantum optics: the Avalanche Photodiode (APD) and the Homodyne Detection (HD). 

\subsection{Non-classicality of a measurement}
There are different signatures for the non-classicality of states such as the negativity in particular quasi-probability distributions or the contextuality of hidden variable models trying to describe the measurement results \cite{Bell1966, Kochen1967}. 
These two notions are in fact equivalent, as it was recently shown in Ref. \cite{Spekkens2008}.

The \textit{non-classicality of a measurement} corresponds to the \textit{non-classicality of its pre-measurement state}, for which such a notion is well-established with different signatures.

We illustrate the relevance of such a correspondance for optical detectors in the conditional preparation of non-classical states of light.
In such experiments \cite{Lvovsky2001, Ourjoumtsev2006}, we generally reconstruct from experimental data the Wigner representation of the conditioned state (\ref{cond_state}) given by:
\begin{equation}
\label{Wigner_cond_state}
\mathcal{W}_{\mathrm{cond}}^{[n]}\left(x,p\right)=\mathcal{N}\int\,dx'dp'\,\mathcal{W}_{AB}\left(x,p;x',p'\right)\mathcal{W}_{\mathrm{retr}}^{[n]}\left(x',p'\right),
\end{equation} 
where $\mathcal{W}_{AB}$ and $\mathcal{W}_{\mathrm{retr}}^{[n]}$ are respectively the Wigner representations for the resource $\hat{\rho}_{AB}$ and the pre-measurement state $\hat{\rho}_{B,\mathrm{retr}}^{[n]}$ retrodicted from the expected result 'n'. $\mathcal{N}$ is a positive normalizing constant of $\mathcal{W}_{\mathrm{cond}}^{[n]}$.
When the resource has a non-negative representation as gaussian states, the necessary condition for preparing a non-classical state $\hat{\rho}_{\mathrm{cond}}^{[n]}$ is to perform a non-classical measurement, in the sense of a non-positive Wigner representation $\mathcal{W}_{\mathrm{retr}}^{[n]}$.

In pratice, we use for such a task an Avalanche Photodiode (APD) which is a single-photon detector with only two responses, called 'off' and 'on'. 
The response 'off' corresponds to the imperfect detection of zero photon (see for example \cite{Barnett1998}):
\begin{equation}
\label{OFF_POVM}
\hat{\Pi}_{\mathrm{off}}\left(\eta,\nu\right)=e^{-\nu}\,\sum_{n=0}^{\infty}\left(1-\eta\right)^{n}\vert n\rangle\langle n\vert,
\end {equation}
in which $\eta$ and $\nu$ are respectively the detection efficiency and the mean number of dark counts. 
On the other side, the Wigner representation \cite{Wigner1932} of the POVM element $\hat{\Pi}_{\mathrm{on}}=\hat{1}-\hat{\Pi}_{\mathrm{off}}$ is given by:
\begin{equation}
\label{wigner_ON}
\mathcal{W}_{\mathrm{on}}\left(x,p\right)=\frac{1}{2\pi}-\frac{e^{-\nu}}{\pi}
\sum_{m=0}^{\infty}\,\left(\eta -1\right)^{m}\,e^{-\left( x^{2}+p^{2}\right)}L_{m}\left[2\left(x^{2}+p^{2}\right)\right].
\end{equation}
where $L_{m}\left(x\right)=e^{x}\partial_{x}^{m}\left(x^{m}e^{-x}\right)/m!$ are the Laguerre polynomials.

\begin{figure}[h!]
\begin{center}
\includegraphics[scale=0.3]{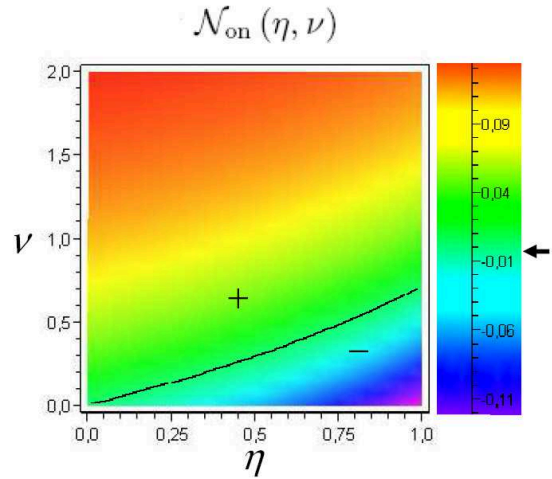}
\caption{\label{negativity_APD} (color online) Evolution of the non-classicality of a measurement performed by an APD, displaying the result 'on', with its detection efficiency $\eta$ and the mean number of dark counts $\nu$. The black line corresponds to a negativity $\mathcal{N}_{\mathrm{on}}\left(\eta,\nu\right)=\mathcal{W}_{\mathrm{on}}\left(0,0\right)=0$.}
\end{center}
\end{figure}

When we read the result 'on', the non-classicality of the measurement performed by the APD can be measured by the \textit{negativity} of the Wigner representation of its pre-measurement state. 
To avoid some mathematical difficulties, we choose for the negativity $\mathcal{N}_{\mathrm{on}}\left(\eta,\nu\right)=\mathcal{W}_{\mathrm{on}}\left(0,0\right)$, since we have $\mathrm{Tr}\lbrace\hat{\Pi}_{\mathrm{on}}\rbrace >0$.
The evolution of this signature of non-classicality, under the noise influence, is pictured on Fig. \ref{negativity_APD}.

Thus, this signature stays negative for a mean number of dark counts such that $\nu<-\ln {\left(1-\eta/2\right)}$. We can easily check this for a given APD, by simply measuring its detection efficiency $\eta$ and its dark noise $\nu$. 
Moreover, we can compare such a signature of non-classicality for different "on/off" photon detectors, by simply plotting their points $\left(\eta,\nu\right)$ on Fig. \ref{negativity_APD}.

\subsection{Projectivity} 
An \textit{ideal} measurement checks a \textit{simple proposition} corresponding to a projector $\hat{\Pi}_{n}=\vert\psi_{n}\rangle\langle\psi_{n}\vert$ in the Hilbert space.
However, in more realistic situations, a measuring device is characterized by POVM elements which are not at all projectors. 
We have actually an evaluation of the \textit{projectivity of a measurement} with the purity $\pi_{n}$ of its pre-measurement state:
\begin{equation}
\label{projectivity}
\pi_{n}=\mathrm{Tr}\left[\left(\hat{\rho}_{\mathrm{retr}}^{[n]}\right)^{2}\right],
\end{equation}
such that $1/D <\pi_{n}\leq 1$.

When the pre-measurement state is a pure quantum state with $\pi_{n}=1$, the measurement performed by the apparatus is \textit{projective} for the response 'n', but it may be \textit{non-ideal}. 
Indeed, the POVM element corresponding to such a projectivity $\left(\pi_{n}=1\right)$ is given by
\begin{equation}
\hat{\Pi}_{n}=\eta_{n}\,\vert\psi_{n}\rangle\langle\psi_{n}\vert
\end{equation}
where $\eta_{n}=\mathrm{Tr}\lbrace\hat{\Pi}_{n}\rbrace$ can be viewed as the detection efficiency of the state $\vert\psi_{n}\rangle$, by using the predictive approach in which the predictive probability $\mathrm{Pr}\left(n\vert\psi_{n}\right)=\eta_{n}$.
We clearly see on this property the complementarity between the predictive and retrodictive approaches in quantum physics.

For an APD, the projectivity is only relevant for the measurement corresponding to the response 'off' which is characterized in the ideal case $\left(\eta=1\right)$ by the vacuum state $\vert 0\rangle$.
Indeed, the pre-measurement state retrodicted from the result 'off' is
\begin{equation}
\label{premeasstate_OFF}
\hat{\rho}_{\mathrm{retr}}^{[\mathrm{off}]}\left(\eta,\nu\right)=\eta\sum_{n=0}^{\infty}\,\left(1-\eta\right)^{n}\vert n\rangle\langle n\vert,
\end{equation}
and the projectivity of this measurement is given by: 
\begin{equation}
\label{projectivity_OFF}
\pi_{\mathrm{off}}\left(\eta,\nu\right)=\frac{\eta}{2-\eta}.
\end{equation}
Contrary to the POVM element (\ref{OFF_POVM}), we can see that this state and its properties do not depend on the dark noise $\nu$. We can easily understand this since the dark counts correspond to results 'on'.
Thus, when we have the result 'off', the properties of the performed measurement only depends on the detection efficiency $\eta$. If we have a quasi-ideal measurement $\eta\simeq 1$, we are sure of the pre-measurement state even for a dark noise arbitrarely large, which may be interesting for conditional protocols based on such measurements.

\subsection{Fidelity with a projective measurement}
We define this fidelity as the overlap between the pre-measurement state $\hat{\rho}_{\mathrm{retr}}^{[n]}$ retrodicted from a certain result 'n' and a target state $\vert\psi_{\mathrm{tar}}\rangle$, in which we would like checking the system before its interaction with the apparatus.
Such a fidelity \cite{Jozsa1994} can be written as
\begin{equation}
\label{fidelity}
\mathcal{F}_{n}\left(\psi_{\mathrm{tar}}\right)=\langle\psi_{\mathrm{tar}}\vert\hat{\rho}_{\mathrm{retr}}^{[n]}\vert\psi_{\mathrm{tar}}\rangle.
\end{equation}
With the retrodictive approach, we have an interesting interpretation for this overlap. This is nothing else than the retrodictive probability (\ref{retrodictive_proba}) of preparing the system in the target state $\vert\psi_{\mathrm{tar}}\rangle$, before the measurement process giving the result 'n':
\begin{equation}
\label{fidelity_proba}
\mathcal{F}_{n}\left(\psi_{\mathrm{tar}}\right)=\mathrm{Pr}\left(\psi_{\mathrm{tar}}\vert n\right)=\mathrm{Tr}\lbrace\hat{\rho}_{\mathrm{retr}}^{[n]}\hat{\Theta}_{\mathrm{tar}}\rbrace.
\end{equation}
The proposition operator (\ref{proposition_operator}) about the state of the system, just after its preparation, is
\begin{equation}
\label{target_proposition}
\hat{\Theta}_{\mathrm{tar}}=\vert\psi_{\mathrm{tar}}\rangle\langle\psi_{\mathrm{tar}}\vert.
\end{equation}
We note that the marginal probability $\mathcal{P}_{\mathrm{tar}}=\mathrm{Pr}\left(\psi_{\mathrm{tar}}\right)$ of preparing the system in this target state is $1/D$ in this particular case. 
It ensures a maximum mixture of pure states constituting a basis of the Hilbert space. This basis is composed of the target state $\vert\psi_{\mathrm{tar}}\rangle$ and $\left(D-1\right)$ other pure states.
Thus, for having an exhaustive description of the system, we need at least one other proposition:
\begin{equation}
\label{off-target_proposition}
\hat{\Theta}_{\,\overline{\mathrm{tar}}}=\hat{1}-\hat{\Theta}_{\mathrm{tar}}.
\end{equation}
It corresponds to the preparation of the system in the state $\hat{\rho}_{\,\overline{\mathrm{tar}}}=\hat{\Theta}_{\,\overline{\mathrm{tar}}}/\mathrm{Tr}\lbrace\hat{\Theta}_{\,\overline{\mathrm{tar}}}\rbrace$ with the probability $\mathcal{P}_{\,\overline{\mathrm{tar}}}=\left(D-1\right)/D$.
When the measurement giving the result 'n' is sufficiently \textit{faithful} $\mathcal{F}_{n}\left(\psi_{\mathrm{tar}}\right)\simeq 1$, the most probable state in which the system was prepared before its interaction with the apparatus is this target state $\vert\psi_{\mathrm{tar}}\rangle$.

\begin{figure}[h!]
\begin{center}
\includegraphics[scale=0.37]{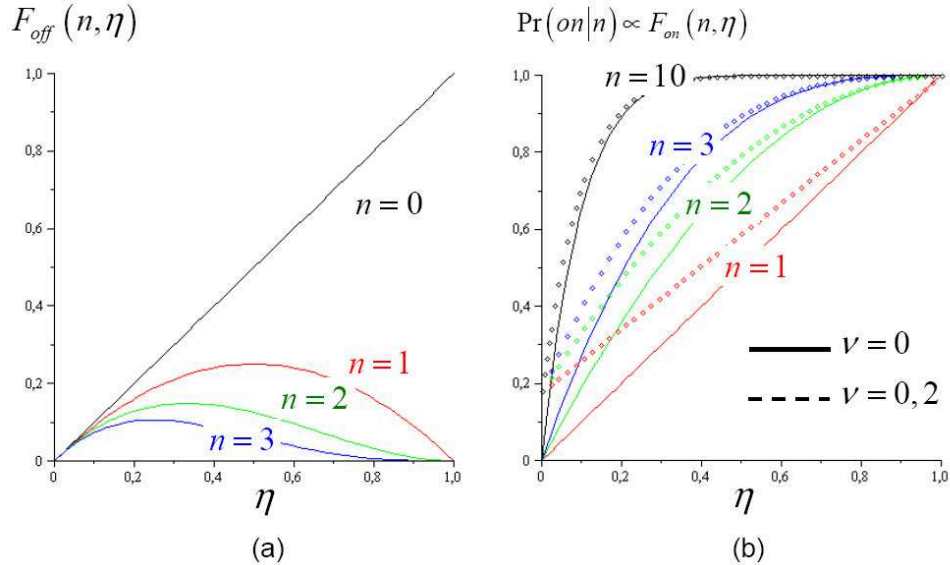} 
\end{center}
\caption{\label{fidelity_APD} (color online) Fidelities between the measurements performed by an APD and PNRDs : (a) Evolution of the fidelity, between the measurement corresponding to the result 'off' and projective measurements of $n$ photons, with the quantum efficiency $\eta$. (b) Evolution of the predictive probability $\mathrm{Pr}\left(\mathrm{on}\vert n\right)$ with the quantum efficiency $\eta$ and the photon-number $n$, for two mean numbers of dark counts $\nu$.}
\end{figure}

We now illustrate this notion for the measurements performed by an APD.
For the measurement corresponding to the result 'off', the fidelity with an ideal photon number-resolved detection (PNRD) of $n$ photons is given by,
\begin{equation}
\label{fidelity_OFF}
\mathcal{F}_{\mathrm{off}}\left(n,\eta\right)=\mathrm{Pr}\left(n\vert\mathrm{off}\right)=\eta\left(1-\eta\right)^{n}.
\end{equation}
The evolution of this fidelity with the detection efficiency $\eta$ is plotted for different photon numbers $n$ on Fig. \ref{fidelity_APD} (a).
As we can expect it for the response 'off' of an efficient APD, the only PNRD which is the most faithful to the measurement performed by the APD corresponds to the photon number $n=0$.
In terms of retrodictive probabilities (\ref{fidelity_proba}), this means that the most probable state in which the system was prepared, before the measurement giving the result 'off', is the vacuum state.
For a low efficiency $\eta\rightarrow 0$, all the photon-number states become equally probable. 
We can understand this since the measurement is not at all projective $\left(\pi_{\mathrm{off}}\simeq 0\right)$ for such efficiencies.

For the measurement corresponding to the result 'on', the situation is more delicate since the POVM element corresponding to this result leads to divergences in the retrodiction of the pre-measurement state.
Indeed, we can only determine an asymptotic equivalent for this pre-measurement state:
\begin{equation}
\label{premeasurement_state_ON}
\hat{\rho}_{\mathrm{retr}}^{[\mathrm{on}]}\left(\eta,\nu\right)\underset{D\rightarrow +\infty}\thicksim \frac{1}{D}\sum_{n=0}^{D}\left[1-e^{-\nu}\left(1-\eta\right)^{n}\right]\vert n\rangle\langle n\vert,
\end{equation}
where $D$ is in fact the dimension of the Hilbert space.
The fidelity between this measurement and a PNRD of $n$ photons is given by
\begin{equation}
\label{fidelity_ON}
\mathcal{F}_{\mathrm{on}}\left(n,\eta,\nu\right)=\mathrm{Pr}\left(n\vert\mathrm{on}\right)
\underset{D\rightarrow +\infty}\thicksim\frac{1}{D}\,\mathrm{Pr}\left(\mathrm{on}\vert n\right),
\end{equation}
in which we recognize the predictive probability (\ref{conditional_proba}) of having the result 'on' when the light was prepared in the photon-number state $\vert n\rangle$:
\begin{equation}
\label{pred_proba_ON}
\mathrm{Pr}\left(\mathrm{on}\vert n\right)=1-e^{-\nu}\left(1-\eta\right)^{n}.
\end{equation}
Thus, the behavior of the fidelity $\mathcal{F}_{\mathrm{on}}\left(n,\eta,\nu\right)$ is the same as that of the predictive probability (\ref{pred_proba_ON}).
Its evolution with the photon number $n$ and the detection efficiency $\eta$ is pictured on Fig. \ref{fidelity_APD} (b).
For a given efficiency, the large photon-number states are the most probable preparations leading to the result 'on', with or without dark noise. We can also see the effect of this one on Fig. \ref{fidelity_APD} (b), which is mainly a non-zero probability for a totally inefficient detector $(\eta=0)$. 
This probability is the same for all the photon-number states. We understand this since the APD has actually no interaction with the signal field. 

\subsection{Non-Gaussian character of a measurement}
The \textit{non-Gaussian character of a measurement} is essential in many quantum information protocols such as entanglement purification \cite{Eisert2002}.
This property can be measured by the \textit{non-Gaussian character of its pre-measurement state}.
We can compute for instance its "non-Gaussianty" (non-G) \cite{nonG2008} defined by the relative Von Neumann entropy between this pre-measurement state and its reference gaussian state, which is the gaussian state with the same covariance matrix. The non-G is therefore equal to zero only for gaussian states.

Then, there is an interesting result based on Hudson-Piquet's theorem \cite{Hudson1974}. 
It states that any pure state, characterized by a non-negative Wigner representation \cite{Wigner1932}, is a gaussian state with a gaussian distribution for its Wigner representation.
We have here an important link between the Gaussian character of a measurement and its projectivity : When a measurement is \textit{projective} with a \textit{non-negative} Wigner representation for its pre-measurement state, then this measurement is \textit{a Gaussian measurement}.

In order to illustrate such a Gaussian character, we determine the pre-measurement states retrodicted from results of a measuring device widely used in quantum optics : the Homodyne Detection (HD).
Indeed, the inefficient HD of the value $x_{I}$ for the quadrature observable $\hat{x}_{\phi}$ \cite{quadrature_def} is characterized by the following POVM element \cite{Banaszek1997}:
\begin{equation}
\label{POVM_HD}
\hat{\Pi}_{\mathrm{HD}}\left(x_{I};\phi\right) = \frac{1}{\sqrt{\pi\left(1-\eta\right)}}\,\exp{\left[-\frac{\left(x_{I}-\sqrt{\eta}\hat{x}_{\phi}\right)^{2}}{1-\eta}\right]},
\end{equation}
in which $\eta$ is the detection efficiency linked to quantum efficiencies of photodiodes used in such a device.
The Wigner representation of this POVM element is then given by:
\begin{equation}
\label{Wigner_HD}
\mathcal{W}_{\mathrm{HD}}\left(x_{\phi},p_{\phi}\right)=
\frac{1}{\sqrt{8\pi^{3}\eta\,\sigma_{x}^{2}\left(\eta\right)}}\,
\exp{\left[-\frac{\left(x_{\phi}-x_{I}/\sqrt{\eta}\right)^{2}}{2\,\sigma_{x}^{2}\left(\eta\right)}\right]},
\end{equation}
with the variance $\sigma_{x}^{2}\left(\eta\right)=\frac{1-\eta}{2\eta}$.

\begin{figure}[h!]
\begin{center}
\includegraphics[scale=0.38]{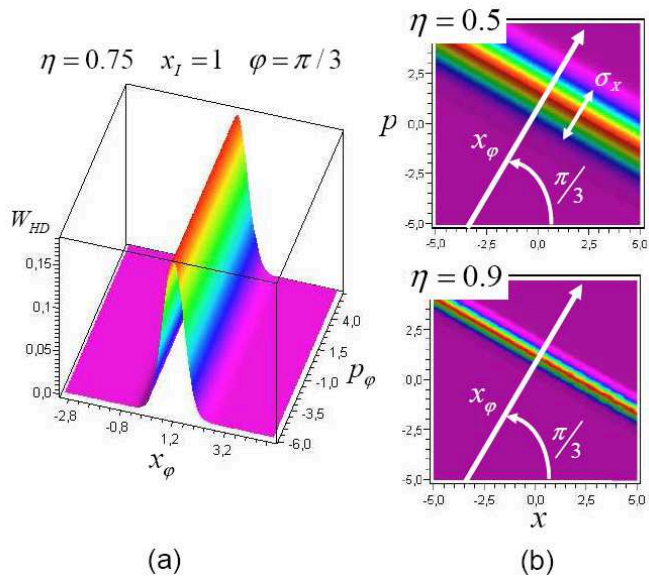} 
\end{center}
\caption{\label{wigner_HD} (color online) (a) Wigner representation of the POVM element $\hat{\Pi}_{\mathrm{HD}}\left(x_{I};\phi\right)$ for $x_{I}=1$ and $\eta=0.75$ in the phase-space $\left(x_{\phi},p_{\phi}\right)$. (b) Evolution of this Wigner representation with the detection efficiency $\eta$ in the phase-space (x,p).}
\end{figure}

The Gaussian character of this measurement is obvious as pictured on Fig. \ref{wigner_HD}, in which we can distinguish the evolution of the variance $\sigma_{x}^{2}\left(\eta\right)$ for different values of the detection efficiency $\eta$.

However, the retrodiction of pre-measurement states needs some mathematical precautions since the trace of POVM elements (\ref{POVM_HD}) diverges. 
After taking a judicious equivalent for the Wigner representation (\ref{Wigner_HD}), we obtain for the pre-measurement state retrodicted from the inefficient HD of the value $x_{I}$:
\begin{equation}
\label{retrostateHD}
\mathcal{W}_{\mathrm{retr}}^{[x_{I};\phi]}\left(x_{\phi},p_{\phi}\right)\underset{e_{n}\rightarrow +\infty}\thicksim
\frac{1}{\pi\sqrt{1+e_{n}\,s\left(\eta\right)}}
\exp {\left[-\frac{\left(x_{\phi}-x_{I}/\sqrt{\eta}\right)^{2}}{s\left(\eta\right)}-\frac{p_{\phi}^{2}}{1/s\left(\eta\right)+e_{n}}\right]}
\end{equation}
We recognize a squeezed coherent state with an amplitude $\alpha_{I}=x_{I}/\sqrt{2\eta}$ and a squeezing parameter $s\left(\eta\right)=\left(1-\eta\right)/\eta$, but with a very large excess noise $e_{n}$ following the conjugate quadratue $p_{\phi}$.

We can therefore say that an HD performs non-classical measurements in the same sense as squeezed states are non-classical states. Indeed, the squeezing of the pre-measurement state (\ref{retrostateHD}) is ensured for a detection efficiency such that $\eta > 50\,\%$. 

In the continuous-wave (cw) regime, the detection efficiency $\eta$ can currently reach values around $98\%$  \cite{Wakui2007}, corresponding to a really impressive squeezing level $s_{\mathrm{dB}}\simeq -17\,\mathrm{dB}$. 
At our knowledge, one of best levels recorded in the cw regime \cite{Schnabel2008} is $-10\,\mathrm{dB}$, corresponding to an efficiency around $90\%$.
Obviously, the pre-measurement state retrodicted from such a detection has a lower purity also corresponding to the projectivity of the HD.

\section{Conclusion}
In this paper, we have shown that the retrodiction of pre-measurement states, in addition to the usual predictive approach, leads to a complete study of the measurement apparatus as an implementation of tests checking propositions about the system interacting with it.
These propositions simply correspond to POVM elements describing the behavior of its responses in the predictive approach. When a certain proposition is checked, the pre-measurement state retrodicted from this one highlights the behavior of the apparatus in terms of states, which is the main language of quantum physics.
We have also proposed an experimental procedure, essentially based on the retrodictive approach, allowing the direct tomography of pre-measurement states for making such a translation.
We hope that this experimental procedure will be realized for probing each response of a measurement device. 
These results, or those from the predictive version \cite{QDT1}, could then be interpreted in the framework of our work, as we have illustrated it on the APD and the HD.


\begin{thebibliography}{99}
\bibitem{quadrature_def} \begin{small}The quadrature observable is defined by $\hat{x}_{\phi}=\left(e^{-i\phi}\hat{a}+e^{i\phi}\hat{a}^{\dagger}\right)/\sqrt{2}$, where $\hat{a}$ (resp. $\hat{a}^{\dagger}$) is the annihilation (resp. creation) operator of a photon satisfying the commutation relation $\left[\hat{a};\hat{a}^{\dagger}\right]=\hat{1}.$
The conjugate quadrature observable $\hat{p}_{\phi}$ is defined by: $\hat{p}_{\phi}=\hat{x}_{\phi+\pi/2}$, and we have the canonical commutation relation $\left[\hat{x}_{\phi};\hat{p}_{\phi}\right]=i\hat{1}$.
The eigen-states of the observable $\hat{x}_{\phi}$ are noted $\vert x,\phi\rangle$, and are a resolution of the Hilbert space, $\int\,dx\,\vert x,\phi\rangle\langle x,\phi\vert =\hat{1}$.
Each eigen-state can be obtained by a rotation in the phase-space: $\vert x,\phi\rangle = \hat{R}_{\phi}^{\dagger}\vert x, 0\rangle$, where $\hat{R}_{\phi}=e^{-i\phi\hat{a}^{\dagger}\hat{a}}$ is the phase-space rotation operator with $\langle x,0\vert x',0\rangle =\delta\left(x-x'\right)$.\end{small}
\end{thebibliography}
\end{document}